\documentclass[12pt]{article}

\usepackage{graphicx,float,latexsym,psfrag,subfigure,fullpage,color}
\usepackage{amssymb,amsmath,amsthm}
\usepackage{bm,multirow}

\usepackage[font=footnotesize,format=plain,labelfont=bf]{caption}

\usepackage{setspace}
\usepackage[square,sort,comma,numbers]{natbib}

\usepackage{textcomp}
\usepackage{gensymb}  

\usepackage{pgfplots}
\usetikzlibrary{spy,backgrounds}
\pgfplotsset{compat=newest}

\usepackage[colorinlistoftodos, color=blue!20!white, bordercolor=gray,
textsize=tiny,textwidth=1.1in]{todonotes}
\definecolor{bluep}{rgb}{0.0,0.6,1}

\newcounter{marknumber}
\pgfplotsset{
    error bars/every nth mark/.style={
        /pgfplots/error bars/draw error bar/.prefix code={
            \pgfmathtruncatemacro\marknumbercheck{mod(floor(\themarknumber/2),#1)}
            \ifnum\marknumbercheck=0
            \else
                \begin{scope}[opacity=0]
            \fi
        },
        /pgfplots/error bars/draw error bar/.append code={
            \ifnum\marknumbercheck=0
            \else
                \end{scope}
            \fi
            \stepcounter{marknumber}    
        }
    }
}

\usepackage{abstract}

\begin{document}

\begin{center}
{\Large \textbf{A mechanism for damage spread by ultrasound-induced linear defects in soft materials}}

\bigskip
\textbf{Pooya Movahed\footnote{pooyam@illinois.edu}}
 
\medskip

Department of Mechanical Science and Engineering\\
University of Illinois at Urbana--Champaign\\
1206 West Green Street, Urbana, IL 61801, USA \\

\medskip

\textbf{Jonathan B. Freund}
 
\medskip

Department of Mechanical Science and Engineering\\
Department of Aerospace Engineering \\
University of Illinois at Urbana--Champaign\\
1206 West Green Street, Urbana, IL 61801, USA \\

\medskip

\today
\medskip

\end{center}

\begin{abstract}
High-intensity ultrasound excites pre-existing bubbles in tissue-like material, and the subsequent bubble
activity may lead to damage. To investigate such damage mechanisms, agar tissue-mimicking
phantoms were subjected to multiple pressure wave bursts of the kind being considered specifically 
for burst wave lithotripsy (BWL). A notable feature of these experiments was the formation of distinct linear defects, as might expected for a crack. We propose a physical mechanism for how a single bubble,
excited by ultrasound, may evolve to from such features. It entails fracture, bubble stability and rectified mass diffusion. These observed linear defects are a potential mechanism for spread of ultrasound-induced bubbles and the subsequent damage.
\end{abstract}

\section{Introduction}

\begin{figure}[!t]
\centering
\begin{tikzpicture}
 \node[inner sep=0pt] (50) at (0,0)
    {\includegraphics[width=.4\textwidth]{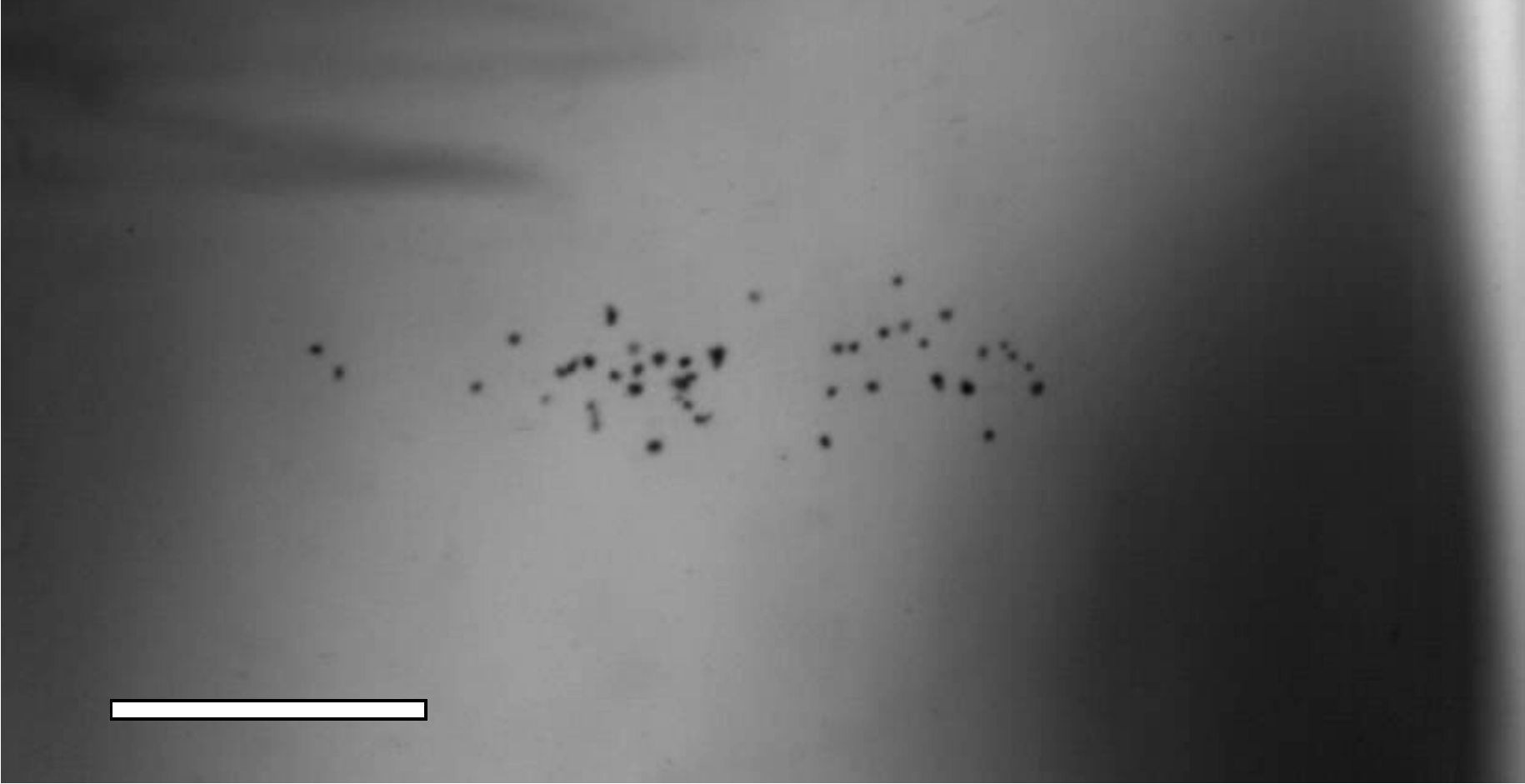}};
  \node[color=white] at (-1.8,-0.9)  {\footnotesize{1\,cm}}; 
 \node[color=white] at (-1.8,1.1) {\footnotesize{(a) pulse 1}};    

 \node[inner sep=0pt] (50) at (5.7,0)
    {\includegraphics[width=.4\textwidth]{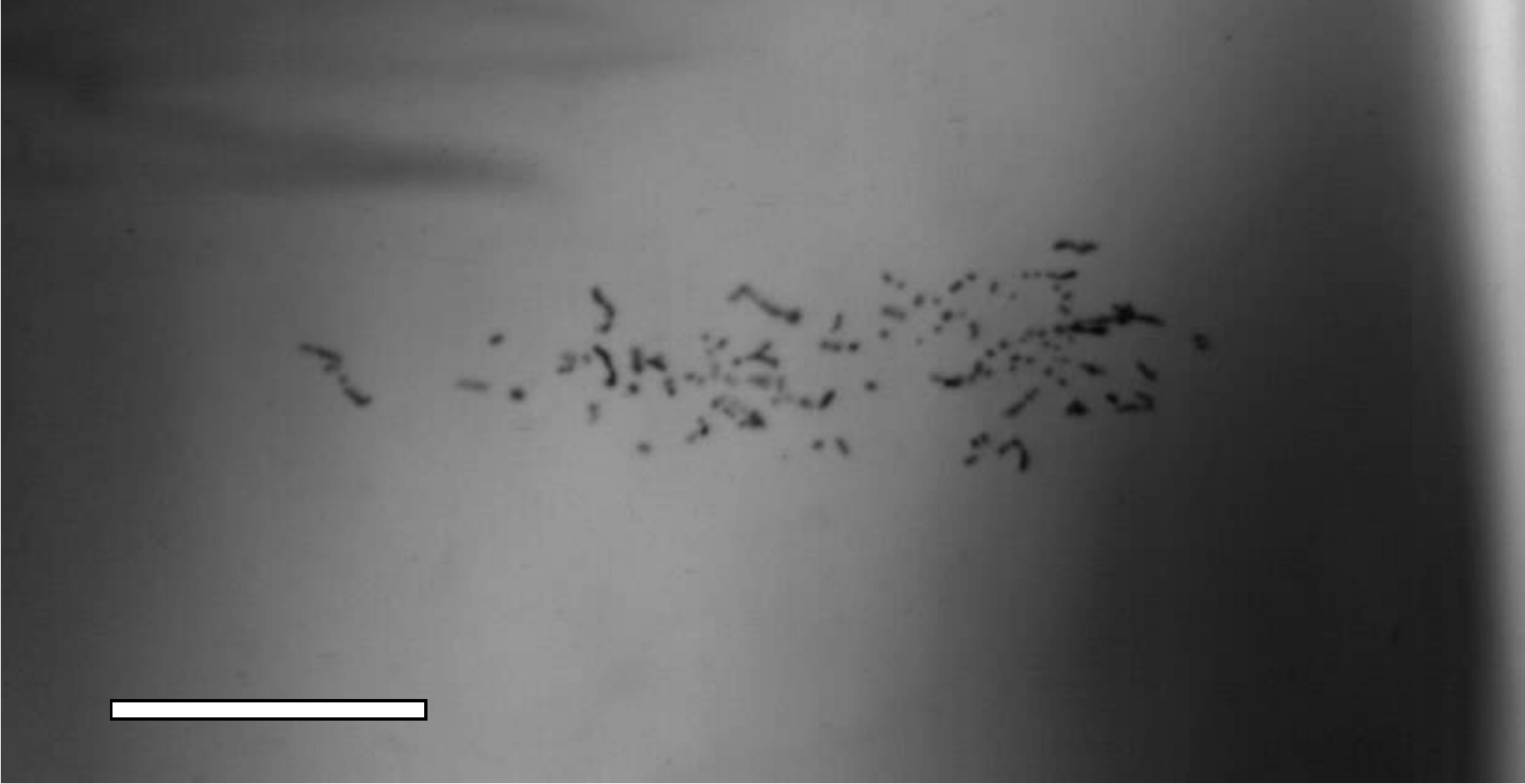}};
  \node[color=white] at (3.9,-0.9)  {\footnotesize{1\,cm}};
  \node[color=white] at (4,1.1) {\footnotesize{(b) pulse 80}};

 \node[inner sep=0pt] (50) at (0,-3) 
    {\includegraphics[width=.4\textwidth]{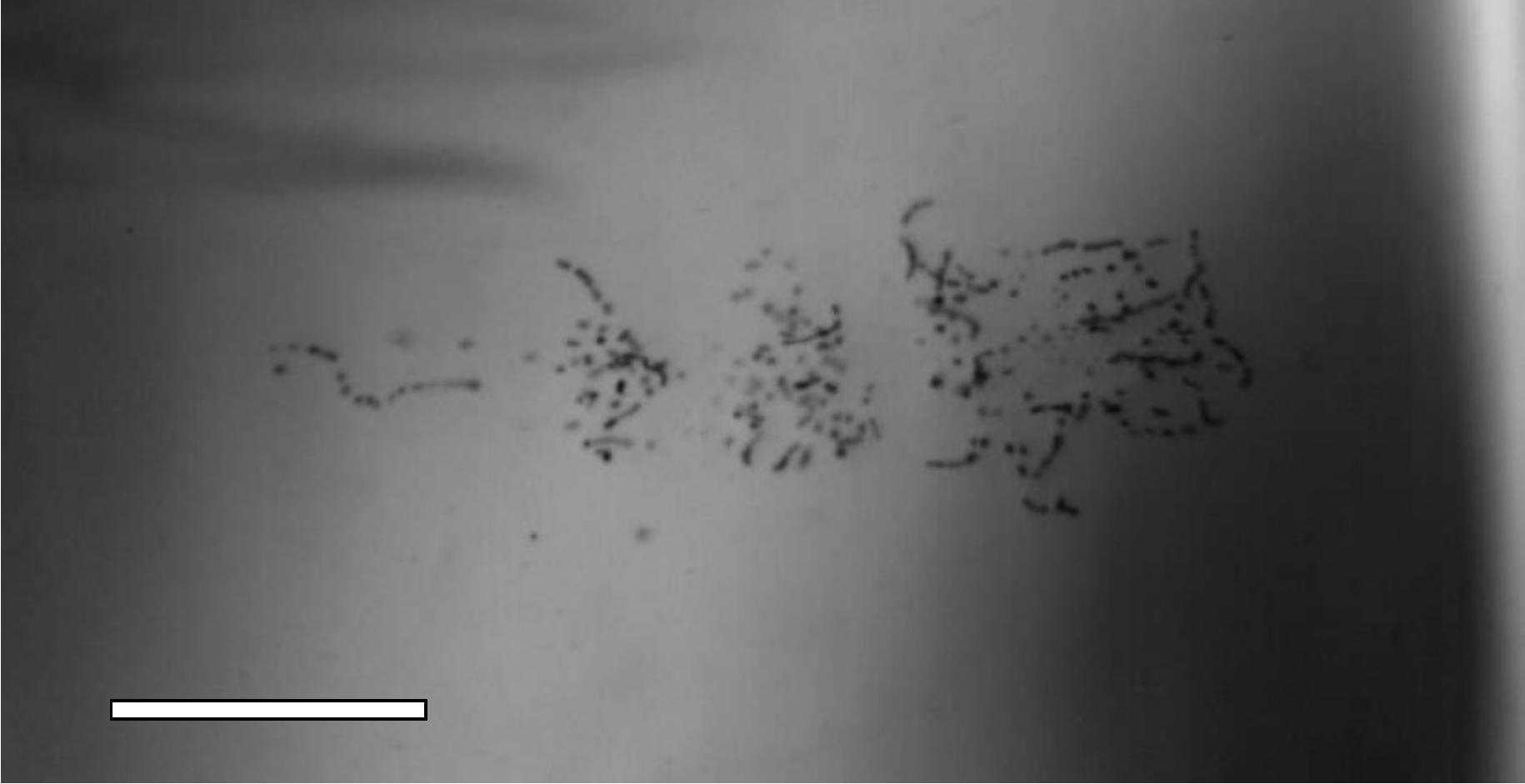}};
  \node[color=white] at (-1.8,-3.9) {\footnotesize{1\,cm}};
  \node[color=white] at (-1.6,-1.9) {\footnotesize{(c) pulse 500}};     

 \node[inner sep=0pt] (50) at (5.7,-3)
    {\includegraphics[width=.4\textwidth]{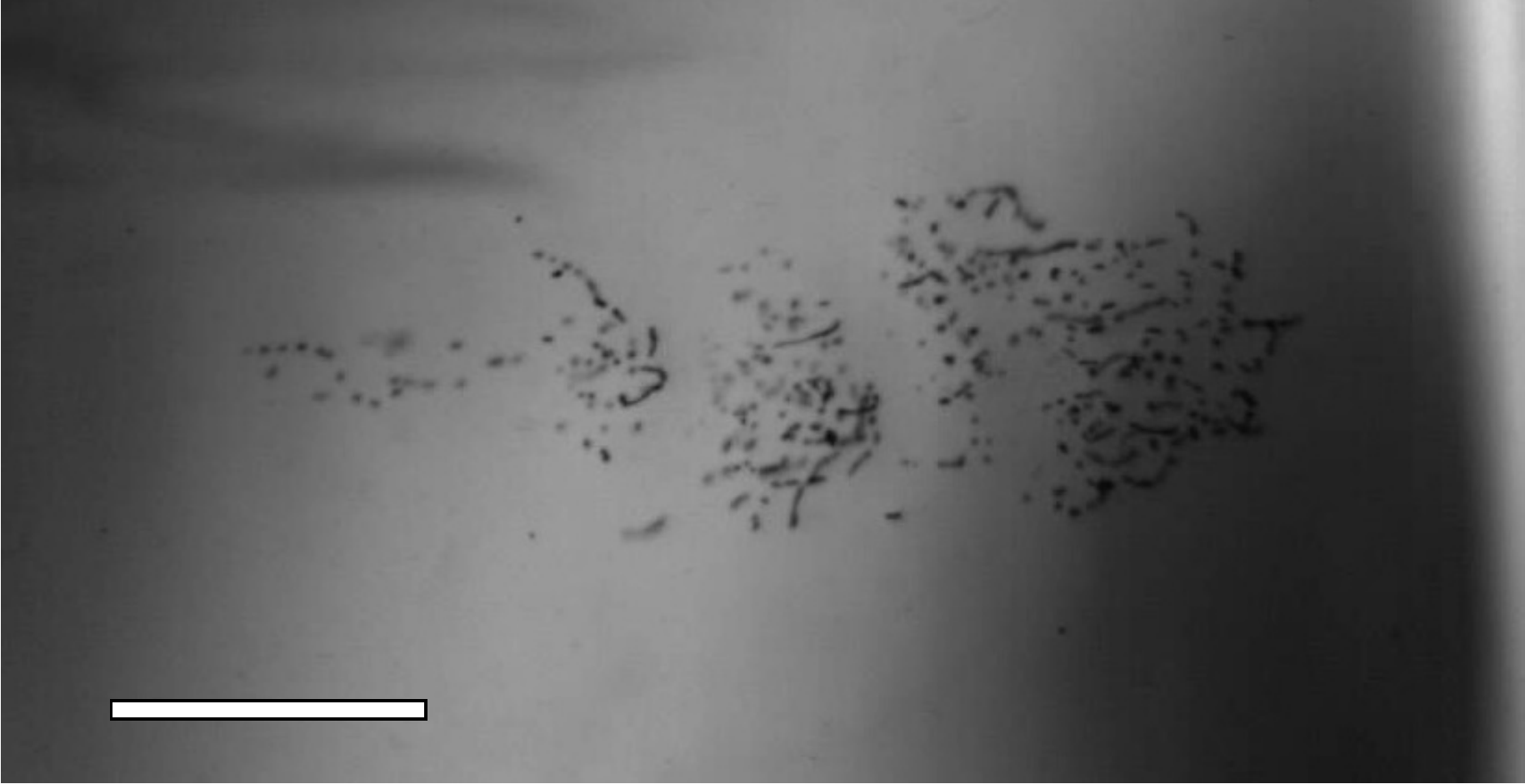}};
  \node[color=white] at (3.9,-3.9)  {\footnotesize{1\,cm}};
  \node[color=white] at (4.2,-1.9) {\footnotesize{(d) pulse 1000}}; 
    \end{tikzpicture}
    
    \caption{Bubbles in 2\% agar gels. The
      region shown is $4.42\, \mathrm{cm}\times 2.27\, \mathrm{cm}$.
      The BWL pulse has 20 nominal cycles with base frequency
      $335\, \mathrm{kHz}$, peak positive pressure amplitude of $10.09\, \mathrm{MPa}$, peak negative pressure amplitude of
      $-7.51\, \mathrm{MPa}$ and repeats at $200\, \mathrm{Hz}$.
      Bubbles and defects are shown 10 cycles into the labeled pulse number. 
}
	\label{fig:agar_1}
\end{figure}
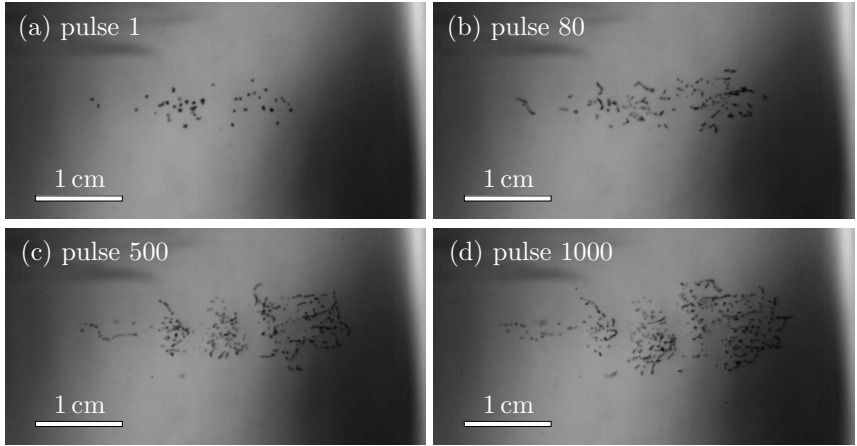

\begin{figure}[!t]
\centering
 \begin{tikzpicture}
 \node[inner sep=0pt] (1) at (0,0)
 {\includegraphics[width=\textwidth]{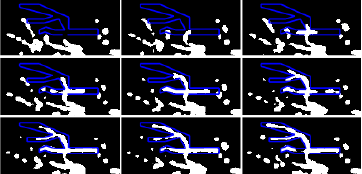}};
  \end{tikzpicture}
  \caption{Tunnel formation and branching in 2\% agar gel. Blue lines are added on top of the black \& white shadowgraphic figures to facilitate the observation of linear defects growth. Frames correspond to every sixth burst.}
\label{fig:tunnel}
\end{figure}

To investigate damage mechanisms induced by excited bubbles by ultrasound, agar tissue-mimicking
phantoms were subjected to 1000 BWL bursts and observed with a high-speed camera \cite{pooya2017} as shown in figure \ref{fig:agar_1}.
A particularly notable features of the images is
the appearance of what seem to be lines of bubbles, which one can
imagine sitting along a crack or some similar some sort of line
defect. An example of their formation is shown in stages in
figure~\ref{fig:tunnel}.  The bubble regions spread from pre-existing
bubble sites into linear structures with branches.  During its
evolution, these features appear to become transiently disconnected in places,
 though these reconnect during the subsequent
pulses. We investigate these features in the context of 
mechanisms by which ultrasound-induced 
damage may progress once a cloud has formed.

Similar features, there called tunnels, have been observed by Williams \& Miller \cite{williams2003} for a
0.78\,MHz ultrasound wave with 4\,MPa amplitude focused on an
agarose phantom for 10\,seconds. More tunnels were observed with inclusion of
Optison\circledR\ ultrasound contrast agents in the phantom to
enhance the number of nucleation sites, and it was suggested that the
formed tunnels could be tracked back to a single nucleation site, typically the embedded cavitation nucleus. The number of tunnels also increased with higher-intensity pulses and longer exposure times. Multiple tunnel branchings
were reported, and these were associated with bubble fusion and/or fragmentation within the tunnels. The possible presence of multiple bubbles within these tunnels was consistent
with observed bubble motion.
Numerous attempts to section these gels and drain the content of these tunnels were unsuccessful, and
consequently it was postulated
that these tunnels were filled with a relatively high viscosity liquid
or partially re-solidified gel. 

In another study, Casket et al.~\cite{caskey2009} investigated vessel wall disruption by insonified microbubbles in the context of diagnostic imaging and drug delivery. They observed tunnel formation by insonified microbubbles in vessels within agarose hydrogels, and the vessel disruption was quantified by measuring the length and width of the tunnel. The tunnel width corresponded to the maximum expansion of the bubbles, and the tunnel length was correlated with the pulse duration. The peak pulse pressure threshold for disruption decreased for larger bubble concentrations, and the tunnel formation was reported only on the distal side of the vessel relative to the ultrasound transducer. While these tunnels formed with a range of orientations relative to the incoming ultrasound pulse, they have a distinct tendency to form in parallel to the pulse propagation.

Our primary goal is to propose a specific physical mechanism for the development of linear defects such as 
seen in figure~\ref{fig:tunnel}. It is based upon the formation of cracks at large strains in soft materials \cite{hashemnejad2015,grzelka2017,o2018experimental},  bubble fragmentation due to shape instabilities \cite{hao1999}, and rectified diffusion for oscillating bubbles subjected to high-intensity ultrasound \cite{sapozhnikov2002}. This is  important as a potential  mechanism for damage spread in ultrasound medical treatments. While we report results here in the context of BWL, this mechanism is also potentially relevant to other treatments, such as HIFU \cite{watkin1996} and histotripsy \cite{macoskey2018using}. We introduce the overall mechanism in the next section, explaining how sub-mechanisms of crack initiation, bubble fragmentation, and rectified diffusion potentially lead a single bubble to evolve to form and grow a linear defect.

\section{Formation of linear defects\label{sec:tunnel}}

.

We first consider a single spherical bubble that oscillates in response to the incident pressure pulses. We will discuss its evolution in the following stages to form a linear defect similar to the one shown in figure~\ref{fig:tunnel}:
\begin{itemize}
\item Stage I: initial crack formation,
\item Stage II: formation of new nucleation sites due to shape instabilities, and
\item Stage III: formation of linear defects.
\end{itemize}
In the subsequent subsections, we describe these stages in detail.

\begin{figure}[!t]
\centering
\resizebox{\textwidth}{!}{
 \begin{tikzpicture}
         \draw [blue,line width=4pt,solid] (0,0) circle [radius=2] node [] {}; 
         \draw [blue,line width=3pt,solid] (0,0) circle [radius=0.3] node [] {}; 
      \fill[black] (1.3,1.5) --(4,4) -- (1.5,1.3) -- cycle;
	\draw [color=blue,->,line width=2pt,dashed] (0.3,0)--(2,0);
	\draw [color=blue,->,line width=2pt,dashed] (-0.21,-0.21)--(-1.4,-1.4);
	\draw [color=blue,->,line width=2pt,dashed] (-0.21,0.21)--(-1.4,1.4);
	\draw [color=black,->,line width=2pt] (1.4,1.4)--(0.0,2.8);
	\draw [color=black,->,line width=2pt] (1.4,1.4)--(2.8,0.0);
         \node[color=black!99] at (3.6,2.7) {\Large{crack}};	
         \node[color=black!99] at (3.8,-0.4) {\Large{max tension}};	
         \node[color=blue] at (0,-2.4) {\Large{bubble expansion}};
          \node[color=black] at (-0.6,4) {\Large{crack initiation}};
          \node[color=black] at (0.0,0.0) {A};	
           \node[color=black] at (-5,2) {\Large{Stage I}};
           
         \draw [blue,line width=4pt,solid] (0,-8) circle [radius=2] node [] {}; 
         \draw [blue,line width=3pt,solid] (0,-8) circle [radius=0.3] node [] {}; 
         \fill[black] (1.3,-6.5) --(4,-4) -- (1.5,-6.7) -- cycle;
	\draw [color=blue,->,line width=2pt,dashed] (0.3,-8)--(2,-8);
	\draw [color=blue,->,line width=2pt,dashed] (-0.21,-8.21)--(-1.4,-9.4);
	\draw [color=blue,->,line width=2pt,dashed] (-0.21,-7.79)--(-1.4,-6.6);	
	\draw [color=black,->,line width=2pt] (1.4,-6.6)--(0.0,-5.2);
	\draw [color=black,->,line width=2pt] (1.4,-6.6)--(2.8,-8.0);	
         \node[color=black!99] at (3.6,-5.3) {\Large{crack}};	
         \node[color=black!99] at (3.8,-8.4) {\Large{max tension}};	
         \node[color=blue] at (0,-10.4) {\Large{bubble expansion}};
         \node[color=black] at (-0.6,-4) {\Large{crack initiation}};
         \node[color=black] at (0.0,-8) {A};	        
         \draw [color=black,->,line width=2pt] (3.5,-8)--(7.3,-8.0);
         \draw [blue,line width=3pt,solid] (8,-8) circle [radius=0.3] node [] {}; 
         \node[color=black] at (8,-8.0) {A};	
         \draw [blue,line width=3pt,solid] (10,-6) circle [radius=0.3] node [] {}; 
         \node[color=black] at (10,-6.0) {B};
         \draw [color=blue,-,line width=3pt] (8.21,-7.79)--(9.79,-6.21);
         \node[color=black] at (9,-4) {\Large{bubble fragmentation}};
         \node[color=blue] at (9,-8.7) {\Large{bubble collapse}};       
         \node[color=black] at (-5,-6) {\Large{Stage II}};
         \draw [blue,line width=4pt,solid] (8,-18) circle [radius=0.3] node [] {}; 
         \draw [blue,line width=3pt,solid] (10,-16) circle [radius=0.3] node [] {}; 
         \draw[color=blue,line width=2pt] (8,-16) arc (90:360:2) ;
         \draw[color=blue,line width=2pt] (10,-18) arc (-90:180:2) ;

       \fill[black] (12,-15.9) --(14,-16) -- (12,-16.1) -- cycle;
       \draw [color=blue,->,line width=2pt,dashed] (10.21,-15.79)--(11.4,-14.6);
	\draw [color=blue,-,line width=4pt,solid] (9.71,-16.21)--(8.21,-17.79);
	\draw [color=blue,->,line width=2pt,dashed] (10.21,-16.21)--(11.4,-17.4);
	\draw [color=blue,->,line width=2pt,dashed] (9.79,-15.79)--(8.6,-14.6);
	\draw [color=blue,->,line width=2pt,dashed] (7.79,-17.79)--(6.6,-16.6);
	\draw [color=blue,->,line width=2pt,dashed] (7.79,-18.21)--(6.6,-19.4);
	\draw [color=blue,->,line width=2pt,dashed] (8.21,-18.21)--(9.4,-19.4);	
         \node[color=black!99] at (13,-15.5) {\Large{crack}};	
          \node[color=black!99] at (10,-16) {B};	
          \node[color=black!99] at (8,-18) {A};
          \node[color=blue] at (12.7,-18.5) {\Large{bubble expansion}};          
          \draw [blue,line width=4pt,solid] (-1,-18) circle [radius=0.3] node [] {}; 
          \draw [blue,line width=3pt,solid] (1,-16) circle [radius=0.3] node [] {}; 
          \draw [color=blue,-,line width=4pt,solid] (0.79,-16.21)--(-0.79,-17.79);
          \node[color=black!99] at (1,-16) {B};	
          \node[color=black!99] at (-1,-18) {A};    
          \node[color=blue] at (0,-18.7) {\Large{bubble collapse}};     
          \node[color=black] at (0,-13.7) {\Large{bubble fragmentation}};
          \node[color=black] at (7,-13.7) {\Large{bubble spread}};
          \draw [color=black,->,line width=2pt,solid] (2,-18)--(5.5,-18);
          \node[color=black] at (-5,-14) {\Large{Stage III}};
         \draw[color=blue,line width=2pt] (-2,-24) arc (90:360:2) ;
         \draw[color=blue,line width=2pt] (0,-26) arc (-90:180:2) ;
         \draw [blue,line width=4pt,solid] (-2,-26) circle [radius=0.3] node [] {}; 
         \draw [blue,line width=3pt,solid] (0,-24) circle [radius=0.3] node [] {}; 
         \fill[black] (2,-23.9) --(4,-24) -- (2,-24.1) -- cycle;
  	 \draw [color=blue,->,line width=2pt,dashed] (0.21,-23.79)--(1.4,-22.6);
         \draw [color=blue,-,line width=4pt,solid] (-0.21,-24.21)--(-1.79,-25.79);
	\draw [color=blue,->,line width=2pt,dashed] (0.21,-24.21)--(1.4,-25.4);
	\draw [color=blue,->,line width=2pt,dashed] (-0.21,-23.79)--(-1.4,-22.6);
	\draw [color=blue,->,line width=2pt,dashed] (-2.21,-25.79)--(-3.4,-24.6);
	\draw [color=blue,->,line width=2pt,dashed] (-2.21,-26.21)--(-3.4,-27.4);
	 \draw [color=blue,->,line width=2pt,dashed] (-1.79,-26.21)--(-0.6,-27.4);	
         \node[color=black!99] at (3,-23.5) {\Large{crack}};	
         \node[color=black!99] at (0,-24) {B};	
         \node[color=black!99] at (-2,-26) {A};
         \node[color=blue] at (2.7,-26.5) {\Large{bubble expansion}};

         \draw[color=blue,line width=2pt] (10,-24) arc (90:360:2);
         \draw[color=blue,line width=2pt] (12,-26) arc (-90:-45:2);
         \draw[color=blue,line width=2pt] (12,-22) arc (90:180:2);
         \draw[color=blue,line width=2pt] (12,-22) arc (90:45:2);
         \draw[color=blue,line width=2pt] (13.4,-25.4) arc (-135:135:2);
         \draw [blue,line width=4pt,solid] (10,-26) circle [radius=0.3] node [] {}; 
         \draw [blue,line width=3pt,solid] (12,-24) circle [radius=0.3] node [] {}; 
         \draw [blue,line width=3pt,solid] (14.8,-24) circle [radius=0.3] node [] {}; 
         \fill[black] (16.8,-23.9) --(18.8,-24) -- (16.8,-24.1) -- cycle;
	\draw [color=blue,->,line width=2pt,dashed] (12.21,-23.79)--(13.4,-22.6);
	\draw [color=blue,-,line width=4pt,solid] (11.79,-24.21)--(10.21,-25.79);
	\draw [color=blue,->,line width=2pt,dashed] (12.21,-24.21)--(13.4,-25.4);
	\draw [color=blue,->,line width=2pt,dashed] (11.79,-23.79)--(10.6,-22.6);
	\draw [color=blue,->,line width=2pt,dashed] (9.79,-25.79)--(8.6,-24.6);
	\draw [color=blue,->,line width=2pt,dashed] (9.79,-26.21)--(8.6,-27.4);
	\draw [color=blue,->,line width=2pt,dashed] (10.21,-26.21)--(11.4,-27.4);
	\draw [color=blue,->,line width=2pt,dashed] (15.1,-24)--(16.8,-24);
	\draw [color=blue,->,line width=2pt,dashed] (14.8,-23.7)--(14.8,-22);
	\draw [color=blue,->,line width=2pt,dashed] (14.8,-24.3)--(14.8,-26);
	\draw [color=blue,-,line width=4pt,solid] (14.5,-24)--(12.3,-24);
	\draw [color=black,->,line width=2pt,solid] (2.3,-25)--(7.5,-25);
         \node[color=black!99] at (18,-23.5) {\Large{crack}};	
         \node[color=black!99] at (12,-24) {B};	
         \node[color=black!99] at (10,-26) {A};
         \node[color=black!99] at (14.8,-24) {C};
         \node[color=blue] at (14.7,-26.5) {\Large{bubble expansion}};    

\end{tikzpicture}
}
  \caption{Schematic of evolution of a single bubble to a linear defect by ultrasound bursts.}
\label{fig:schematic}
\end{figure}
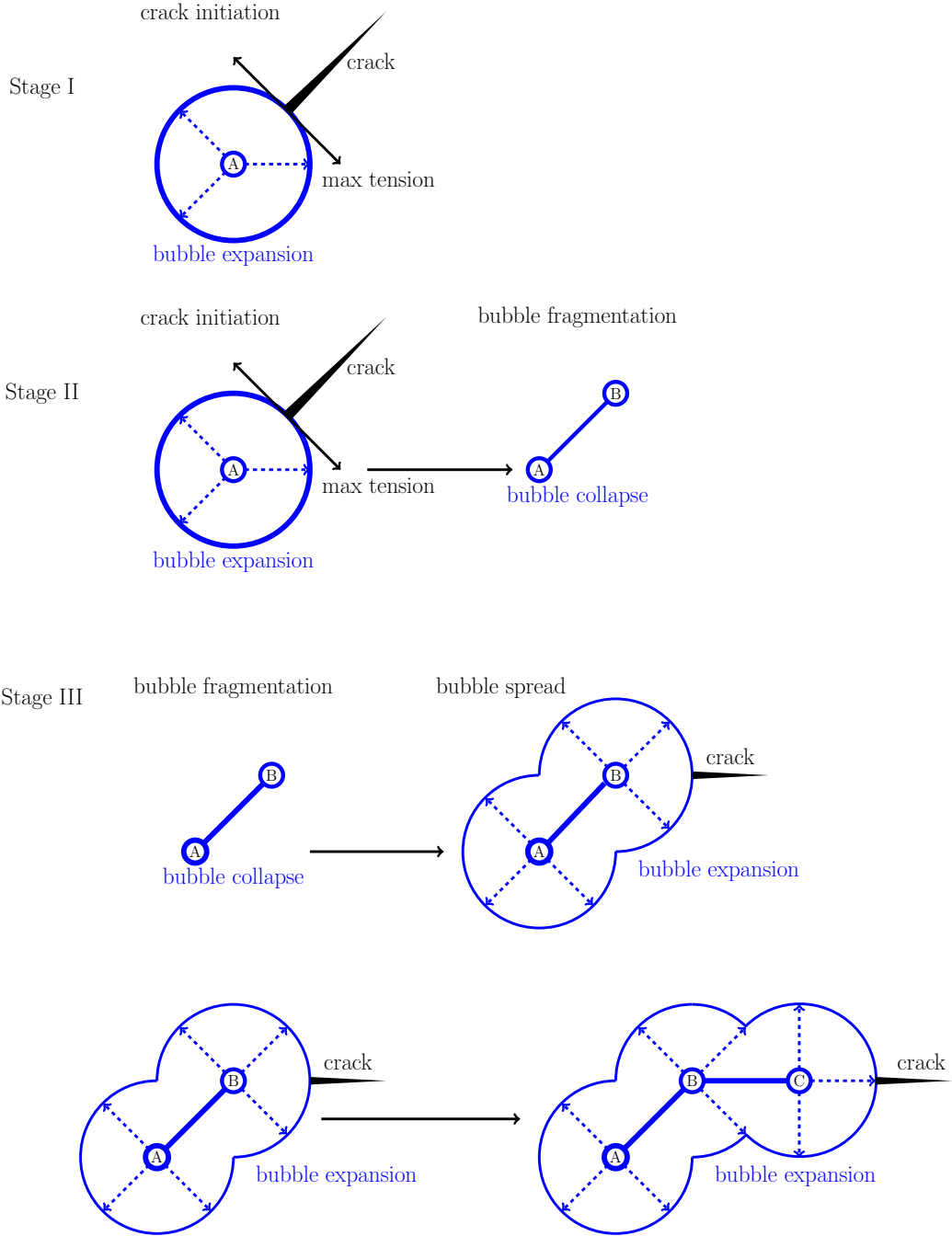

\subsection{Stage I: initial crack formation}

There is extensive experimental support \cite{barrangou2006,pooya2016b,catheline2004,pooya2017,nayar2012} for the Fung model \cite{fung1979,fung2013biomechanics} affording a reasonable quantitative description of the mechanical behavior of agar gel. In this framework, agar gel is represented by a strain energy density 
\begin{equation}
U=\frac{\eta}{2\alpha} \exp^{\alpha (I_1-3)},
\end{equation}
where $\eta$ and $\alpha=1$ \cite{pooya2016b} are material properties, and $I_1$ is the first principal invariant of the left Cauchy--Green tensor. For small strains, $\eta$ is the low-strain shear modulus,
and was measured using microindentometry to be $54\,\mathrm{kPa}$ for our particular agar gels  \cite{pooya2017}.

Assuming spherical symmetry, the corresponding contribution to the
elastic stress tensor at the bubble surface $R$ is  \cite{Johnsen2015,pooya2016b,bader2018}
\begin{equation}
\sigma_{e,rr}=\frac{2\eta}{3}\left[ \bigg(\frac{R_o}{R}\bigg)^4-\bigg(\frac{R}{R_o}\bigg)^2\right] \exp^{\alpha \left(I_1-3\right)}-p,
\label{eq:stress1}
\end{equation}
\begin{equation}
\sigma_{e,\theta \theta}=\sigma_{e,\phi \phi}=\frac{\eta}{3}  \left[4\bigg(\frac{R}{R_o}\bigg)^2- \bigg(\frac{R_o}{R}\bigg)^4\right] \exp^{\alpha \left(I_1-3\right)}-p,
\label{eq:stress2}
\end{equation}
where  $p$ is the hydrostatic pressure, and 
\begin{equation}
I_1=\bigg(\frac{R_o}{R}\bigg)^4+2\bigg(\frac{R}{R_o}\bigg)^2.
\end{equation}

This approach was used in a quantitative model for bubble dynamics with damage mechanisms based on fracture and fatigue \cite{pooya2016b}, and the model successfully reproduced experimental observations \cite{pooya2017}. Here we use it to justify the likelihood of crack formation, such as in the schematics of figure~\ref{fig:schematic}.  For such large strains, any such constitutive model, even for so flexible a material as agar, is expected to satisfy standard criterion for fracture. Particularly, for a bubble with initial radius in the range $0.01 \mu \mathrm{m} \le R_o \le 1\mu \mathrm{m}$ in agar gel, which has $10^2\,\mathrm{Pa}\le \eta \le 10 ^6\, \mathrm{Pa}$, the Griffith theorem \cite{griffith1921} predicts a specific fracture stretch ratio $\lambda_f$ in the range $1.45 \le \lambda_f \le 3.05$ \cite{pooya2016b}.

The maximum observed bubble radius in our experiments is $\sim230\,\mu \mathrm{m}$~\cite{pooya2017}. Consequently, assuming a submicron initial bubble radius, this criterion suggests that the stretch ratio at the bubble surface indeed exceeds the fracture threshold predicted by the Griffith theorem. As the bubble expands ($R/R_o \gg1$), the stresses are such that  $\tau_{\theta \theta}, \tau_{\phi \phi}\gg \tau_{rr}$ based on (\ref{eq:stress1}) and (\ref{eq:stress2}).
Since the theory of fracture dynamics predicts crack initiation in the plane of maximum stress,
the initial crack is expected to form in the $r$-direction as shown in figure~\ref{fig:schematic}, although cracks may form at different angles if there is any particular hydrogel weakness at that site. Since the camera pixel resolution of $58\,\mu  \mathrm{m}$ is much larger than the anticipated width of a typical crack, they are not observable in our particular experiments, though they have been reported elsewhere \cite{hashemnejad2015,grzelka2017,o2018experimental}. 
Movahed \textit{et al.}~\cite{pooya2017} quantified the orientation of bubbles by fitted ellipses, and it was shown that bubbles are most probably oriented parallel to the pulse propagation. A similar tendency for tunnel formation parallel to the wave propagation has also been reported~\cite{caskey2009}.

\subsection{Stage II: formation of new nucleation sites due to shape instabilities}

 \begin{figure}
\centering
\begin{tikzpicture}
	\begin{semilogxaxis}[
		xlabel=$\mu\,(\mathrm{Pa\,s})$,
		ylabel=$R_{\text{crit}}\,(\mathrm{nm})$,
		ymax=500,
		ymin=0,
		ytick={0,150,300,450},
		xtick={0.001,0.01,0.1},
		xmin=0.001,
		xmax=0.1,
		width=0.45\textwidth,
		legend style={draw=none,legend pos=south east,legend cell align=left,font=\small}
		]
	\addplot[color=blue,only marks,mark=*] coordinates {
		(0.001,350)
		(0.003,50)
		(0.006,70)
		(0.01,80)
		(0.03,90)
		(0.06,140)
		(0.1,450)
	};
	\end{semilogxaxis}
\end{tikzpicture}
 \caption{The critical bubble radius $R_{\text{crit}}$ vs. viscosity $\mu$ based on the second harmonic amplitude perturbation.}
\label{fig:shape_ins}
\end{figure}
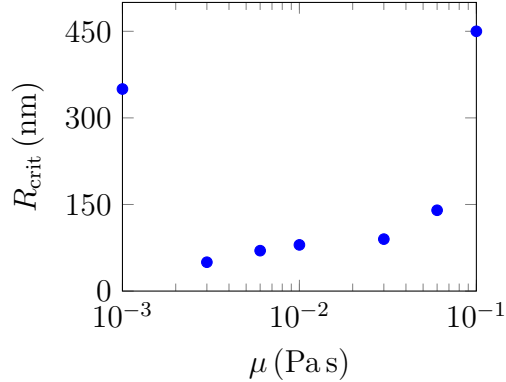

Even an unconfined bubble, in an infinite fluid, is expected to undergo shape instabilities under some conditions, which in turn may cause it to split to multiple bubbles.
We generalize the framework developed by Hao \& Prosperetti~\cite{hao1999} 
to anticipate instabilities of a spherical bubble in a viscous Newtonian fluid. 
In this framework, bubbles with an initial radius of larger than a critical value $R_{\text{crit}}$ are unstable. 
Viscosity measurements for agar gels are unavailable at our specific conditions, which include both high
strains and high strain rates \cite{pooya2016b}. However, agar gels are expected to be highly viscous at these conditions \cite{catheline2004}, and we consider a wide range of viscosity for our analysis. Since the bubble response is expected to 
be dominated by viscosity rather than elasticity for the considered BWL pulse \cite{pooya2016b,pooya2017},
the current approach provides a measure of $R_{\text{crit}}$, which is sufficient
for our development. The current study also motivates the development of
models for shape instabilities in the presence of elasticity \cite{murakami2017}, but it is sufficient here to provide an upper bound for $R_{\text{crit}}$. Elasticity would further restrict the growth of distortions.

The distorted bubble surface is represented  as \cite{plesset1954} 
\begin{equation}
r(t)=R(t)+\sum_{n=1}^{\infty}a_n(t) Y_n,
\end{equation}
where $Y_n$ is a spherical harmonic of degree $n$, $a_n$ is the amplitude of the surface distortion, and $R$ is the evolving mean bubble radius. The time evolution of the distortion amplitude for any $n$ is \cite{hao1999}
\begin{equation}
\begin{split}
\ddot{a}_n+&\bigg[3\frac{\dot{R}}{R}+2(n+2)(2n+1)\frac{\mu}{\rho R^2}\bigg]\dot{a}_n+\\
&(n-1)\bigg[-\frac{\ddot{R}}{R}+(n+1)(n+2)\frac{S}{\rho R^3}+2(n+2)\frac{\mu \dot{R}}{\rho R^3}\bigg]a_n=0,
\end{split}
\label{eq:shape_ins}
\end{equation}
where $\rho=1012.8\,\mathrm{kg/m^{3}}$ is the density, $S=0.073\,\mathrm{N/m}$ is the surface tension, and $\mu$ is the viscosity. We focus on $n=2$, which is the most important for the instability \cite{hao1999}, and the bubble dynamics model (\ref{eq:KM}) in the appendix~\ref{app:bub} is solved simultaneously with (\ref{eq:shape_ins}) for $a_2(0)=0.1\, R_0$ and $\dot{a_2}(0)=0$. The amplitude increases rapidly for bubbles with $R_o \ge R_{\text{crit}}$, while being suppressed for smaller bubbles due to surface tension and viscosity. Figure~\ref{fig:shape_ins} shows $R_{\text{crit}}$ for $\mu=0.001\,\mathrm{Pa\,s}$ up to $0.1\,\mathrm{Pa\,s}$, which represents a range of estimates from the literature \cite{catheline2004,pooya2016b,estrada2018}. Thus we anticipate instability despite the uncertain gel viscosity.

As described in the previous subsection, a crack should develop as the bubble expands during the first stage. 
Given the perturbation to the spherical symmetry of the asymmetric crack, we can anticipate that such instabilities will be seeded and lead to bubble breakdown. With symmetry broken, especially if the initial bubble is less confined in the direction of the new crack, we can anticipate that the bubble splits as it undergoes subsequent oscillations. When this happens in a fluid, it leaves numerous small bubbles in the neighborhood~\cite{hao1999}.  Due to the violent flow in the near neighborhood of bubbles, some of these will move into the crack and provide seed for subsequent bubble growth there, further spreading the crack. In our schematic (figure~\ref{fig:schematic}), we simply depict a parent bubble splitting into two bubbles during the second stage. This process leads to formation of a new nucleation site labeled ``B".

\subsection{Stage III: formation of linear defects}

During the first and second stages, a single bubble causes crack formation during the expansion phase, and bubble fragmentation leads to formation of new nucleation sites, which
will assist further crack growth. In our experiments, the agar phantom is exposed to 1000 pulses. Thus
this process happens continuously, leading to features  growing in time as observed in figure~\ref{fig:tunnel}.

In our proposed mechanism, new bubbles are formed during the second stage, but
these are not expected to be initially large enough to form new cracks when driven to become unstable. The time scale for the growth of these new bubbles is governed by rectified mass diffusion \cite{sapozhnikov2002}, as described in appendix~\ref{sec:rectified}. The rate at which the 
equilibrium bubble radius $R_o$ increases by rectified mass diffusion for our BWL pulses is plotted for a pulse with 10 cycles in figure~\ref{fig:rectified}.
$R_o$ increases from $0.1\,\mu\mathrm{m}$ to $~4\,\mu \mathrm{m}$ during this pulse, which is 
sufficient to make the bubble unstable due to shape instabilities considering the reported values for $R_{\text{crit}}$ in figure~\ref{fig:shape_ins}.

\section{Rate of linear defects growth}

\begin{figure}
\centering
\begin{tikzpicture}
	\begin{axis}[
	 x tick label style={/pgf/number format/.cd,
          scaled x ticks = false,
          set thousands separator={},
          fixed},
		xlabel=$t\,(\mu \mathrm{s})$,
		ylabel=$R_o\,(\mu \mathrm{m})$,
		ymin=0,
		ymax=4,
		xmin=0,
		xtick={0,10,20,30,40},
        		legend style={draw=none,legend pos=south east,legend cell align=left},
		width=0.45\textwidth
		]
	\addplot+ [color=red, ,line width=1pt,solid,mark=none, error bars/.cd,y dir=both,y explicit] 
	table [x expr=\thisrowno{0}/335*1000, y expr=\thisrowno{1}*0.1 ]{Mass_R0n.plt};
	\end{axis}
\end{tikzpicture}
\caption{Temporal evolution of the equilibrium bubble radius $R_{o}$. The viscosity of the surrounding medium is the same
as water ($\mu=0.001\,\mathrm{Pa\,s}$).}
\label{fig:rectified}
\end{figure}
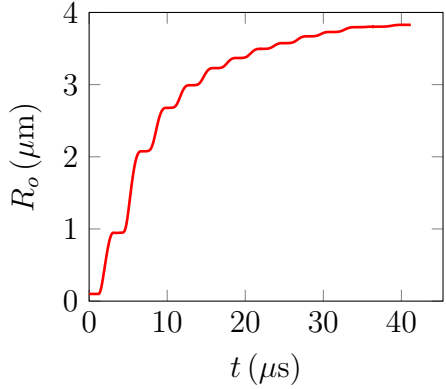

\begin{figure}[!t]
\centering
 \includegraphics[width=0.6\textwidth]{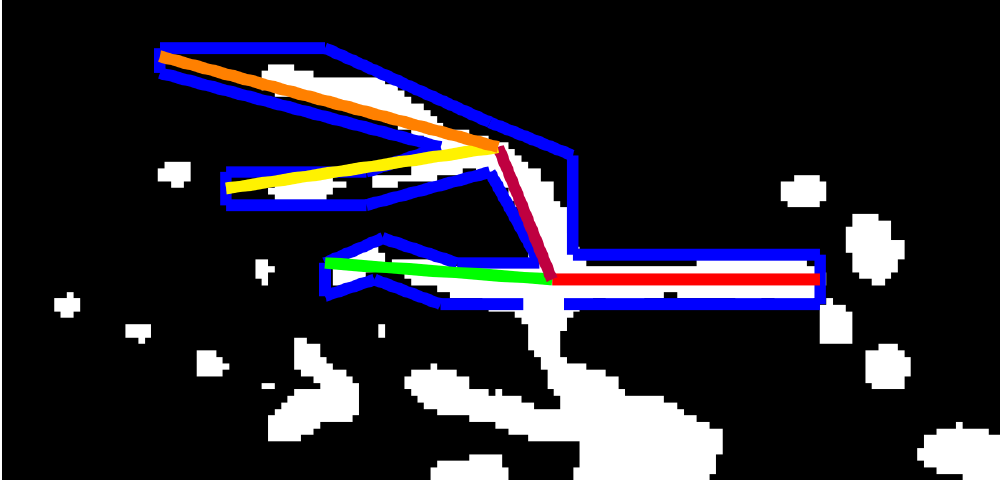}
  \caption{The blue structure becomes filled with bubbles after subsequent BWL pulses. This structure is 
  represented by five lines as shown to facilitate the line defect spread rate measurements, which are summarized in table~\ref{table:growth}.}
\label{fig:growth_schematic}
\end{figure}

To measure rate of growth, the blue structure drawn in figure~\ref{fig:tunnel},
which becomes filled with bubbles after subsequent BWL pulses, is represented by five lines
as shown in figure~\ref{fig:growth_schematic}. The length of each line and the number of pulses
required to fill each line with bubbles are reported in table~\ref{table:growth}. Based on this measurement,
the average line defects growth rate is $79\,\mu\mathrm{m}$  per pulse. Considering that the width of each line is about the maximum observed bubble radius $R_{\textrm{max}}\sim230\,\mu \mathrm{m}$ \cite{pooya2017}, this suggests that the cracks shown in figure~\ref{fig:schematic} grow at the rate of $R_{\textrm{max}}/3$ per pulse in our experiments. Considering that each pulse consists of 20 cycles, an important question is why the growth rate is limited to $R_{\textrm{max}}/3$ per pulse. While crack formation and bubble fragmentation may happen during each cycle, rectified diffusion is the dominant mechanism dictating the growth rate as it requires multiple cycles to absorb more gas
as shown in figure~\ref{fig:rectified}, which indicates a time scale comparable to observations.

\begin{table}[ht]
\centering
\begin{tabular}{c c c c}
\hline
color & length ($\mu m$) & \# of pulses & growth rate ($\mu m$/pulse) \\ [0.5ex] 
\hline
red &2168& 29 & 74 \\
green&2385& 29& 84 \\
purple&1300&19 & 68 \\
yellow & 2600 & 31 & 84 \\
orange & 3250 & 37 & 88 \\ [1ex]
\hline
average & & & 79
\end{tabular}
\caption{Line defect growth rate per pulse measurements for different lines shown in figure~\ref{fig:growth_schematic}.}
\label{table:growth}
\end{table}

The BWL pulse used in this study has a base frequency of 335\,kHz, and bubbles are expected to oscillate at about the same frequency based on a Rayleigh--Plesset-type bubble dynamics model \cite{pooya2016b,pooya2017}. Due to the high frequency of the bubble oscillations and also the limited camera resolution ($58\,\mu\mathrm{m}$), unfortunately, we cannot provide experimental direct evidence of the presence of cracking. While it is challenging to observe cracks in experiments under
dynamic loading at high strain rates, formation of cracks  has been reported in the literature for soft materials \cite{hashemnejad2015,grzelka2017,o2018experimental}, specifically, in experiments designed to identify material properties under static loading.

\section{Conclusions\label{sec:conclusions}}

Bubble activity in tissue-like material due to high-intensity ultrasound may lead to damage.
To investigate such damage mechanisms, agar tissue-mimicking
phantoms were subjected to multiple pressure wave bursts of the kind being considered specifically 
for burst wave lithotripsy. A three-stage physical mechanism was proposed to describe how a single bubble may evolve
to form linear defects, a notable feature observed in the experiments. They seem to collect along distinct lines, as would expected for a crack.  The growth rate of linear defects is
measured to be $79\,\mu\mathrm{m}\sim R_{\textrm{max}}/3$ per pulse in our experiments, where $R_{\textrm{max}}\sim230\,\mu \mathrm{m}$ \cite{pooya2017}. This rate is in line with estimated rates of rectified mass diffusion and the critical bubble radius for bubble to become unstable due to shape instabilities.

\section*{Acknowledgments}

The authors are grateful for fruitful discussions with W.~Kreider, A.~D.~Maxwell, S.~B.~Hutchens and M.~R.~Bailey. 
This work was supported by NIH NIDDK grant number P01-DK043881.

\appendix

\section{Bubble dynamics model\label{app:bub}}

To study the bubble dynamics for a single spherical bubble, we consider the Keller--Miksis approach, where
the bubble dynamics is governed by \cite{keller1980bubble,plesset1977} 
\begin{equation}
\left(1-\frac{\dot{R}}{c}\right)R\ddot{R}+\frac{3}{2}\left(1-\frac{\dot{R}}{3c}\right)\dot{R}^2=\frac{1}{\rho}\left(1+\frac{\dot{R}}{c}\right)Q+\frac{R}{\rho c} \frac{d}{dt}Q,
\label{eq:KM}
\end{equation}
where $R$ is the bubble radius, $\dot{R}$ and $\ddot{R}$ are velocity and acceleration at the bubble
surface, $c=1480\,\mathrm{m\, s^{-1}}$ is the sound speed and the density of the surrounding medium, and
\begin{equation}
Q=p_B-p_{\infty}-\frac{2S}{R}-\frac{4\mu \dot{R}}{R},
\label{eq:QB}
\end{equation}
where $p_B$ is the internal bubble pressure modeled as a polytropic
gas, and $p_{\infty}\left(t\right)$ is the pressure of the passing BWL pressure pulse \cite{pooya2016b}.

\section{Rectified diffusion\label{sec:rectified}}

To measure the amount of rectified diffusion \cite{sapozhnikov2002} for a single spherical bubble, we consider 
the conservation of gas in the liquid, which is governed by
\begin{equation}
\frac{\partial C_a}{\partial t}+u_r \frac{\partial C_a}{\partial r}=D \nabla^2 C_a,
\label{eq:Ca}
\end{equation}
where $u_r=\frac{R^2\dot{R}}{r^2}$ is the radial velocity, and $C_a=\rho_a/\rho_l$ is the
gas concentration. This equation is solved in the semi-infinite range $R\le r \le \infty$,
and for simplicity, it is mapped onto $0\le x \le 1$, where
\begin{equation}
x=\frac{l}{l+r-R},
\end{equation}
and $l$ is a measure of the mass diffusion length in the liquid. 
The governing equation in the mapped space then becomes
\begin{equation}
\begin{split}
&\frac{\partial C_a}{\partial t}+\frac{x^2}{l}\dot{R}\left(1-\frac{R^2}{(l/x-l+R)^2}\right)\frac{\partial C_a}{\partial x} =\\
& D \frac{x^4}{l^2}\frac{\partial^2 C_a}{\partial x^2}+D\frac{2x^3}{l^2}\frac{\partial C_a}{\partial x}-\frac{2D}{l/x-l+R}\frac{x^2}{l}\frac{\partial C_a}{\partial x}.
\end{split}
\end{equation}
The applied boundary conditions are
\begin{equation}
C_a(r=\infty)=C_a(x=0)=\frac{p_{\infty}}{H}, \qquad C_a(r=R)=C_a(x=1)=\frac{p_i}{H},
\end{equation}
where $H$ is the Henry's law constant, and $p_i$ is the internal bubble pressure.

The first and second derivatives are approximated using fourth-order finite differences on a uniform grid
\begin{equation}
\frac{d C_a}{dx}\bigg|_i=\frac{- C_{a,i+2}+8 C_{a,i+1}-8  C_{a,i-1} +C_{a,i-2}}{12\Delta x},
\end{equation}
and
\begin{equation}
\frac{d^2 C_a}{dx^2}\bigg|_i=\frac{- C_{a,i+2}+16 C_{a,i+1}-10C_{a,i}+16  C_{a,i-1} -C_{a,i-2}}{12\Delta x^2}.
\end{equation}

The mole flux into the bubble (positive) is
\begin{equation}
m''=-4\pi R^2 \frac{1}{l} \frac{d C_a}{dx}.
\label{eq:nflux}
\end{equation}
The number of gas moles inside the bubble gets updated by knowing the mole flux from (\ref{eq:nflux}).
The equilibrium radius also changes with $n$, and by assuming an isothermal process \cite{sapozhnikov2002}, we have
\begin{equation}
\frac{(p_o+2\sigma/R_o) R_o^3}{n_o}=\frac{(p_o+2\sigma/R_{on})R_{on}^3}{n},
\label{eq:Ron}
\end{equation}
where $R_{on}$ is the updated equilibrium radius. The internal gas pressure changes adiabatically
\begin{equation}
p_g R^{3\gamma}=(p_o+2\sigma/R_{on})R_{on}^{3\gamma}.
\label{eq:Radiab}
\end{equation}
(\ref{eq:Ron}) and (\ref{eq:Radiab}) are combined  to obtain \cite{sapozhnikov2002}
\begin{equation}
p_g=\bigg(p_o+\frac{2\sigma}{R_o}\bigg)\frac{n}{n_o}\bigg(\frac{R_o}{R}\bigg)^{3\gamma}\bigg(\frac{R_{on}}{R_o}\bigg)^{3(\gamma-1)}.
\end{equation}

\bibliographystyle{plain}


\end{document}